\documentclass{aa}  

\usepackage{xcolor}
\usepackage{graphicx}
\usepackage{txfonts}
\usepackage{float}
\usepackage{amsmath}
\usepackage{hyperref}

\newcommand\Reym{\mbox{Rm }}  % Peclet number
\newcommand\Bo{\mathbf{\overline{B}}_0}
\newcommand\Bor{{\overline{B}}_{0_x}}
\newcommand\Boy{{\overline{B}}_{0_y}}

\begin{document}
\title{Magnetorotational dynamo chimeras}
\subtitle{The missing link to turbulent accretion disk dynamo models?}
\titlerunning{Magnetorotational dynamo chimeras}

\author{A. Riols\inst{1}\and F. Rincon\inst{2,3}\and
  C. Cossu\inst{4}\and G. Lesur\inst{5,6} \and G. I. Ogilvie\inst{1} \and P-Y. Longaretti\inst{5,6}}
\institute{%
Department of Applied Mathematics and Theoretical Physics, University of Cambridge, Centre for Mathematical Sciences,
Wilberforce Road, Cambridge CB3 0WA, UK 
\email{ar764@cam.ac.uk}
\and
Universit\'e de Toulouse; UPS-OMP, IRAP Toulouse, France 
\email{francois.rincon@irap.omp.eu}
\and
CNRS, IRAP, 14 avenue Edouard Belin, 31400 Toulouse, France 
\and
Institut de M\'ecanique des Fluides de Toulouse (IMFT), CNRS --
Universit\'e de Toulouse, Allée du Pr. Camille Soula, 31400 Toulouse,
France 
\and
CNRS, IPAG, 38000 Grenoble, France 
\and
Univ. Grenoble Alpes; IPAG, 38000 Grenoble, France 
}

\date{\today}

\abstract{%
In Keplerian accretion disks, turbulence and magnetic
fields may be jointly excited through a subcritical dynamo
process involving the magnetorotational instability (MRI). 
High-resolution simulations exhibit a tendency towards
statistical self-organization of MRI dynamo turbulence into  
large-scale cyclic dynamics. Understanding the physical origin 
of these structures, and whether they can be sustained and transport 
angular momentum efficiently in astrophysical conditions, represents 
a significant theoretical challenge. 
The discovery of simple periodic nonlinear MRI dynamo solutions has
recently proven useful in this respect, and has notably served to
highlight the role of turbulent magnetic diffusion in the seeming
decay of the dynamics at low magnetic Prandtl number Pm (magnetic
diffusivity larger than viscosity), a common regime in accretion
disks. The connection between these simple structures and the
statistical organization reported in turbulent simulations remained
elusive, though. Here, we report the numerical discovery 
in moderate aspect ratio Keplerian shearing boxes of new
periodic, incompressible, three-dimensional nonlinear MRI dynamo
solutions with a larger dynamical complexity reminiscent of such
simulations. \textcolor{black}{These ``chimera'' cycles are characterized 
by multiple MRI-unstable dynamical stages, but their basic physical 
principles of self-sustainment are nevertheless identical to those of 
simpler cycles found in azimuthally elongated boxes}. In particular, 
we find that they are not sustained at low Pm either due to subcritical 
turbulent magnetic diffusion. These solutions offer a new perspective 
into the transition from laminar to turbulent instability-driven dynamos, 
and may prove useful to devise improved statistical models of turbulent 
accretion disk dynamos.}


\maketitle
%%%%%%%%%%%%%%%%%%%%%%%%%%%%%%%%%%%%%%%%%%%%%%%%%%

%%%%%%%%%%%%%%%%% BODY OF PAPER %%%%%%%%%%%%%%%%%%

\section{Introduction}
The magnetorotational instability (MRI) is considered one of the
main sources of angular momentum-transporting turbulence in
astrophysical accretion disks. It requires a magnetic
field and a differentially rotating flow whose angular velocity
decreases with distance to the rotation axis
\citep{velikhov59,chandra60,balbus91}. Numerical studies have shown
that in the presence of a constant net vertical magnetic flux, the MRI
acts as a powerful linear instability which amplifies arbitrarily
small perturbations that break down nonlinearly into MHD turbulence
\citep{hawley95,stone96}. The efficiency of the turbulence at
transporting angular momemtum, however, remains a matter of debate and
may depend on dissipative processes. This transport  may notably be
limited in the astrophysically relevant regime of low magnetic
Prandlt number (Pm), where Pm denotes the ratio between the kinematic
viscosity and magnetic diffusivity of the fluid
\citep{lesur07,balbus08,meheut15}.

A distinct but related problem is the origin of the magnetic field that
supports the MRI in such astrophysical systems. In the absence of an
externally imposed field, one possibility is that the field is created
and sustained inside the disk by a turbulent dynamo process. A good
candidate is the so-called subcritical MRI dynamo process, by which 
magnetohydrodynamic (MHD) perturbations excited by the MRI nonlinearly
sustain the large scale field that made the instability possible in
the first place
\citep{hawley95,hawley96,rincon07b,lesur08b,rincon08}. Simulations of
"zero net magnetic flux" configurations have shown that this process 
can self-sustain and lead to MHD turbulence that transports
significant angular momentum. Many simulations of MRI dynamo
turbulence also exhibit self-organized large-scale dynamics
characterised by chaotic reversals of the large-scale magnetic field,
somewhat reminiscent of the "butterfly" diagram of the solar dynamo
\citep{branden95,davis10,simon11,gressel15}. The viability of an MRI
dynamo process in disks remains unclear though, as numerical studies
explicitly taking into account dissipative process suggest that it
cannot be sustained for $\text{Pm}\lesssim 1$ \citep{fromang07b}. The
highest-resolution incompressible simulations to date indicate that no
MRI dynamo can be excited at $\text{Pm}=1$ for magnetic Reynolds
numbers (Rm) as large as $\text{Rm}=45000$ \citep{walker16}, but other
studies have found some dependence of this effect on geometry and
stratification \citep{oishi11,shi16}.

A fundamental physical understanding of the different processes 
at work in the subcritical MRI dynamo appears to be required
to make sense of these different numerical observations.  
Simple three-dimensional nonlinear periodic dynamo solutions 
of the MHD equations \citep[hereafter H11]{Herault2011}
have recently been shown to provide the first germs of nonlinear 
MRI dynamo excitation in the Keplerian differential rotation regime 
in azimuthally elongated shearing box numerical configurations at
transitional kinematic (Re) and magnetic Reynolds numbers 
\citep[hereafter R13]{riols2013}. These solutions represent an interesting
avenue of research to investigate the inner workings and transitional
properties of this dynamo. A recent analysis of the energy budget of a
few such cycles has notably proven useful to pinpoint the possible role of a
``subcritical'' turbulent magnetic diffusion in the seeming 
decay of the dynamics at low Pm \citep[hereafter R15]{riols15}.

One possible caveat with this approach so far, though, has been 
the difficulty to connect these fairly ordered (yet fully
three-dimensional and nonlinear) periodic solutions to
developed turbulent MHD states produced in generic moderate
aspect ratio simulations at larger Re and Rm, and in particular
to statistically self-organized MRI dynamo butterflies.
The aim of this paper is to bridge part of this gap by presenting
several new three-dimensional nonlinear periodic MRI dynamo solutions 
computed in moderate aspect ratio shearing boxes. As we shall see, the
dynamical complexity of these cycles, which we call ``MRI dynamo 
chimeras'' \textcolor{black}{because they involve multiple MRI-unstable 
dynamical stages}, is significantly larger than that of the simpler 
solutions discussed previously, and is reminiscent of the statistical 
\textcolor{black}{behaviour} observed in generic numerical simulations. 
Yet, their sustainment rests on the exact same few linear and nonlinear 
dynamical processes underlying the dynamics of simpler cycles.

The mathematical and numerical frameworks of our study are presented
in Sect.~\ref{framework}, which also provides a discussion of
the effects of changing the aspect ratio in this
problem. Section~\ref{cycles}
documents the dynamical properties of two new pairs of dynamically
complex chimera MRI dynamo cycles computed in moderate aspect ratio
configurations with a
Newton-Krylov algorithm, and shows that their existence is limited
to Pm larger than a few. Section~\ref{budget} extends the magnetic
energy budget analysis of R15 to these structures, and shows that the
enhancement of nonlinear transfers of magnetic energy to small scales
at large Re (``subcritical turbulent magnetic diffusion'') 
prevents the sustainment of these structures at lower Pm, just as 
in the case of simpler cycles. In Sect.~\ref{dynamostat}, we
discuss the semi-statistical nature of these new cycles
and the perspectives that their computation opens for the development
of improved statistical models of Keplerian dynamos, and more
generally instability-driven dynamos. A summary of the main
conclusions and short discussion of how the results may fit 
into the wider astrophysical context concludes the paper.

\section{Equations and numerical framework}
\label{framework}
\subsection{{Model}}
\label{model}
The equations and numerical frameworks have been described
extensively in R13. We only recall the essential points here.
We use the cartesian local shearing sheet
description of differentially rotating flows \citep{goldreich65},
whereby the axisymmetric differential rotation is approximated locally
by a linear shear flow  $\mathbf{U}_x=-Sx\,  \mathbf{e}_y$, and a
uniform rotation rate $\mathbf{\Omega}=\Omega\,  \mathbf{e}_z$, with
$\Omega=2/3 S$ for a Keplerian shear flow. Here $(x,y,z)$ are
respectively the shearwise, streamwise and spanwise directions,
corresponding to the radial, azimuthal and vertical directions in
accretion disks. We refer to the $(x,z)$ projection of vector fields
as their poloidal components and to the $y$ direction as their
toroidal (azimuthal) one. We ignore stratification and compressible effects.
The evolution of the three-dimensional velocity {field} perturbations
$\mathbf{u}$ and magnetic field $\mathbf{B}$ is governed by the
three-dimensional incompressible, dissipative MHD equations:
\begin{equation}
\nabla \cdot\mathbf{u}=0, \quad \nabla\cdot\mathbf{B}=0.
\label{div}
\end{equation}
\begin{equation}
\frac{\partial{\mathbf{u}}}{\partial{t}}-Sx\frac{\partial{\mathbf{u}}}{\partial{y}}+\mathbf{u}\cdot\mathbf{\nabla u} = -2\mathbf{\boldsymbol{\Omega}}\times\mathbf{u}+Su_x\mathbf{e}_y-\mathbf{\nabla}\Pi+\mathbf{B}\cdot\mathbf{\nabla B}+\nu\mathbf{\Delta u},
\label{velocity_eq}
\end{equation} 
\begin{equation}
\frac{\partial{\mathbf{B}}}{\partial{t}}-Sx\frac{\partial{\mathbf{B}}}{\partial{y}}  = -SB_x\mathbf{e}_y+\nabla\times(\mathbf{u}\times\mathbf{B})+\eta\mathbf{\Delta B}.
\label{magnetic_eq}
\end{equation}
where $\Pi$ is the sum of the gas and magnetic pressure. The density is fixed to $\rho=1$. The kinematic and magnetic Reynolds numbers are defined by $\text{Re}=SL^2/\nu$ and
$\text{Rm}=SL^2/\eta$, where $\nu$ and $\eta$ are the constant kinematic viscosity and magnetic diffusivity, $L$ is a typical scale of the
spatial domain and time is measured with respect to
$S^{-1}$. The magnetic Prandtl number is 
$\text{Pm}=\nu/\eta=\text{Rm}/\text{Re}$. $\mathbf{B}$ is expressed in
terms of an alfv\'enic velocity. Both $\mathbf{u}$ and $\mathbf{B}$ are
measured in units of $SL$. 

\subsection{Numerical methods}
\label{numerics}
We use the SNOOPY code \citep{lesur07} to perform direct numerical
simulations (DNS) of Eqs. \eqref{div}-\eqref{magnetic_eq}. This code
provides a spectral implementation of the so-called numerical shearing
box model of the shearing sheet, in a finite domain {of size
$(L_x,L_y,L_z)$, at numerical resolution $(N_x,N_y,N_z)$}. The $x$
and $y$ directions are taken as periodic while shear-periodicity is
imposed in $x$.  As the box is shear-periodic, any numerical solution
can be decomposed into a set of shearing Fourier modes with wavenumbers
\begin{equation}
k_x(t)=-\ell k_{x_0}+mk_{y_0} t~,\quad k_y=m k_{y_0}\quad\text{and}\quad
k_z=n k_{z_0},
\end{equation}
where $k_{x_0}=2\pi/L_x$, $k_{y_0}=2\pi/L_y$, $k_{z_0}=2\pi/L_z$ 
(a detailed description of the spectral decomposition used in
the simulations is provided in R13). A shearing wave is a
non-axisymmetric wave with $m\neq 0$. Because of the shear-periodicity
in $x$, the radial Eulerian wavenumber $k_x$ of a given shearing wave
increases linearly in time ($\ell$ is the corresponding integer
Lagrangian wavenumber). The wave is "leading" when $k_xk_y<0$ and
"trailing" when $k_xk_y>0$.

Nonlinear periodic solutions are computed with the Newton-Krylov
solver PEANUTS (see again R13) interfaced to SNOOPY, and followed in
parameter space using arclength continuation.  We enforce that the
dynamics takes place in a symmetric subspace to facilitate the
analysis (this does not compromise the underlying dynamical
complexity), and notably monitor the axisymmetric MRI-supporting field
$\overline{\mathbf{B}}$ ($(x-y)$ average of $\mathbf{B}$), more
specifically its energetically dominant Fourier mode
$\Bo(z,t)=\Bo(t)\cos(k_{z_0}z)$.

\subsection{Large vs moderate aspect ratio}
\label{aspect_ratio}
In H11, R13 and R15, the nonlinear dynamics was restricted to
elongated boxes in the azimuthal direction with typical azimuthal to
radial aspect ratio $L_y/L_x\sim 30$. The main reason for this choice
was to simplify the dynamics as much as possible and reduce
the number of coexisting cyclic solutions. Indeed, at low Re and Rm
(defined on $L_x\sim L_z$) and $L_y/L_x\gg 1$ the dynamics in the
$(L_x,L_z)$ plane remains quite laminar but the dissipation of
non-axisymmetric structures is sufficiently small that (just a few)
weakly nonlinear three-dimensional structures can be sustained. H11
showed that the 
sustainment of these large-aspect ratio cycles can be understood
in generic physical terms involving shearing effects, the
linear physics of \textcolor{black}{non-axisymmetric MRI-unstable 
perturbations}, and nonlinear feedback mechanisms. One may nevertheless 
argue that large-aspect ratio configurations are somewhat unnatural, and 
indeed the connection with turbulent MRI dynamo simulations conducted
at moderate aspect ratio does not seem straightforward. 

Here, we therefore attempt to extend this ``cycle hunt'' to
moderate aspect ratios, down to $L_y/L_x\sim 4$. 
\textcolor{black}{Because $k_{y_0}$ is larger in this configuration}, 
non-axisymmetric shearing waves are sheared out and
dissipated faster for a given set of Re and Rm (once again defined on
$L_x\sim L_z$). Hence, one has to go to larger 
Re and Rm (typically $\text{Rm}\sim 3000$ for $L_y/L_x\sim
4$) to recover the first germs of sustained (recurrent) MRI dynamo
activity. However, doing so also implies that the dynamics in the $(L_x,L_z)$
plane is more complex than in the large aspect ratio case
because of the larger available number of poloidal dynamical degrees of
freedom. The two main consequences, from the perspective of capturing
nonlinear solutions, are that a larger number of them coexist in the
phase space of the problem for a given Re and Rm, making it more
difficult to converge on any of them, and that resolving them requires
a large numerical resolution. The latter is a significant hurdle in
this context,  as computing a single cyclic solution with a
Newton-Krylov algorithm requires many DNS integrations.
The resolution used in this study is $(48,48,72)$, ensuring
convergence for all parameters considered, but a few results
were also reproduced at the (very computationnally demanding) 
double resolution $(96,96,144)$.

\section{Chimera MRI dynamo cycles}
 \label{cycles}
\subsection{Dynamical structure of the solutions}
\begin{figure*}
\centering
\includegraphics[width=\textwidth]{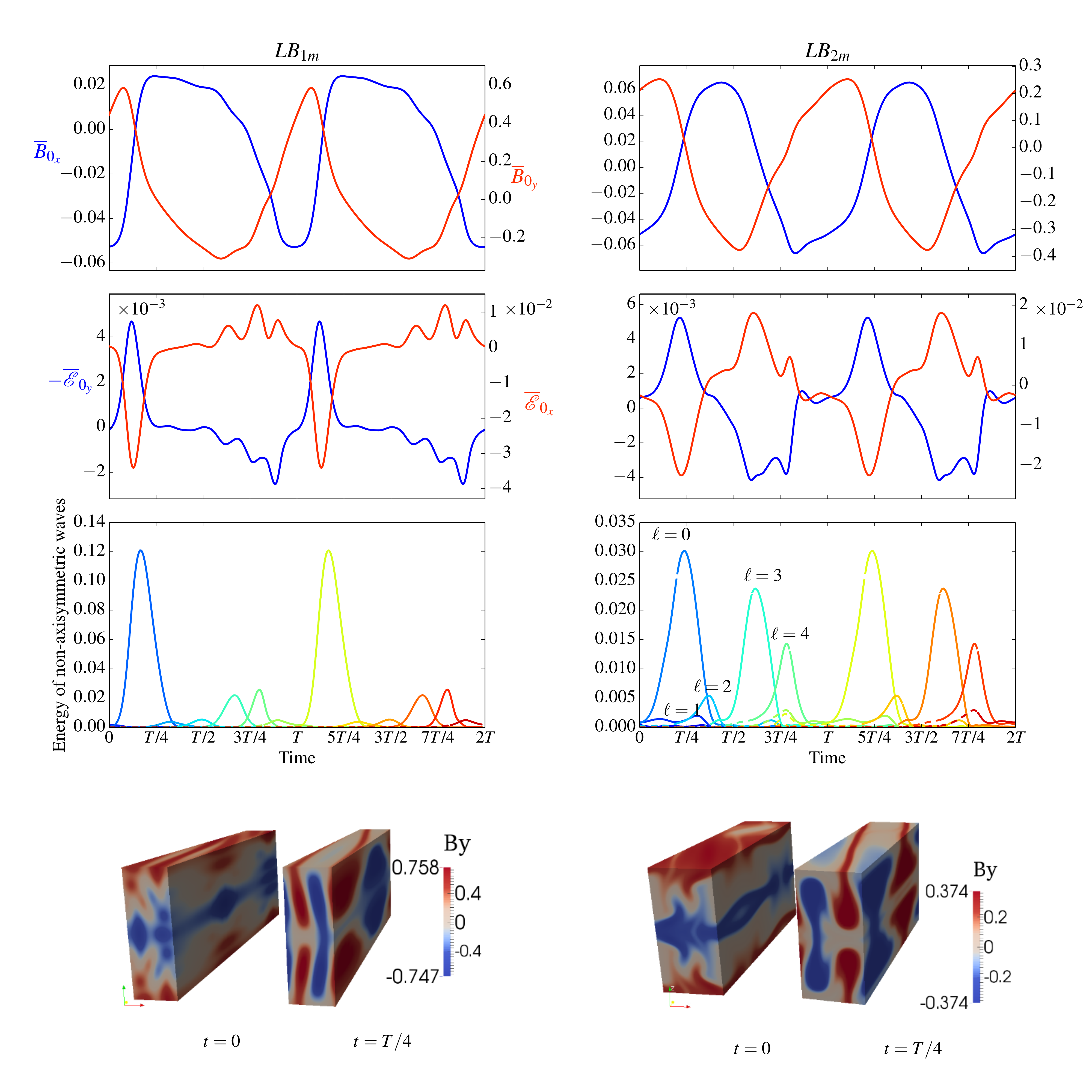}
 \caption{Temporal evolution and dynamical decomposition of the lower
   branches solutions $LB_{1m}$ (left) and $LB_{2m}$
   (right). $LB_{1m}$ is shown for $\text{Re}=100$, $\text{Rm}=900$
   and $LB_{2m}$ for $\text{Re}=908$, $\text{Rm}=3033$. From top to
   bottom, axisymmetric field $\Bo$, axisymmetric EMF
   $(\mathcal{E}_{0x}, -\mathcal{E}_{0y})$ and total energy (integrated over
   all $k_z$) of each shearing wave packet ($k_y=2\pi/L_y$) as a
   function of time (2 cycle periods are represented). The rainbow 
   colors and $\ell$ in the bottom plot represent successive shearing wave
   packets (see Sects.~\ref{framework}-\ref{budget}). The 3D
   visualizations show isosurfaces of $B_y$ at $t=0$
   and $t=T_0/4$, where $T_0$ is the period of the corresponding
 cycle. Videos showing the total $B_y$ and the poloidal velocity streamlines can be found at \url{http://store.maths.cam.ac.uk/ASTRO/ar764/cycle_multish_Ly6.avi} (for $SN_{1m}$) and \url{http://store.maths.cam.ac.uk/ASTRO/ar764/cycle_multish_Ly2.avi} (for $SN_{2m}$).}
\label{fig_cycle_multish}
 \end{figure*}
In order to identify cycles in moderate azimuthal to radial ($L_y/L_x$) aspect ratio shearing box simulations of the MRI dynamo problem, we explored the phase space of the system numerically by constructing turbulence lifetime maps using many different simulations initialised with different initial conditions. This  technique, described in detail in R13, makes it possible to isolate islands of recurrent or chaotic dynamics in phase space. Once an initial condition 
leading to fairly recurrent dynamics was spotted, we used a Newton-Krylov solver described in H11 and R13 to refine the initial condition until a periodic solution is achieved. Using this technique, we managed to capture two pairs of nonlinear periodic solutions. The first
one, labelled $SN_{1m}$, was found in a box of dimensions
$(L_x,L_y,L_z)=(0.7,6,2)$ and has a period $T_{SN_{1m}}=6 \,S^{-1}
L_y/L_x=51.4\,S^{-1}$. The second one, labelled $SN_{2m}$, was found in
a  $(L_x,L_y,L_z)=(0.5,2,1)$ box, and has a slightly shorter period
$T_{SN_{2m}}=5\,S^{-1} L_y/L_x=20\,S^{-1}$.  Each pair of solutions is born out of
a saddle node bifurcation and is therefore composed of a lower branch
cycle and  an upper branch cycle. These solutions are distinct from
the different cycles pairs studied by H11 and R13. The temporal
evolutions of $LB_{1m}$ ($LB_{2m}$), the lower branch of $SN_{1m}$
(respectively $SN_{2m}$), are shown in Fig.~\ref{fig_cycle_multish}.

The physical mechanisms and self-sustaining process underlying the
dynamics of these two cycles are identical to those described by
\cite{rincon07b} and H11. The large scale axisymmetric
component of the field $\Bo$ renders non-axisymmetric
\textcolor{black}{velocity and magnetic perturbations} unstable 
to the MRI in the Keplerian flow. \textcolor{black}{For simplicity, 
we refer to these perturbations as "non-axisymmetric MRI-unstable
shearing wave packets"}\footnote{\textcolor{black}{We use this 
nomenclature because the MRI-unstable perturbations taking part in 
the dynamics documented in both H11 and in the present paper are 
essentially supported by non-axisymmetric $m=1$ shearing Fourier modes. 
However, these perturbations are not reducible to a single Fourier}
\textcolor{black}{mode in $z$, unlike in the uniform toroidal field 
case \citep{balbus92}, because the large-scale axisymmetric 
MRI-supporting "dynamo" field $\Bo$ is non-uniform 
in $z$. More specifically, for the class of solutions considered 
here and in H11, the $z$-dependence of $m=1$ velocity
perturbations is decomposed in $z$ as a sum over Fourier modes
with odd $n$, and that of magnetic perturbations as a sum over
even $n$ (see Appendix of H11). Magnetic perturbations $\vec{b}$ 
with $m=1$ and even $n$ are linearly coupled to velocity 
perturbations $\vec{v}$ with $m=1$ 
and odd $n$ through inductive and Lorentz force terms 
$\Bo\cdot\vec{\nabla}{\vec{u}}$ and $\Bo\cdot\vec{\nabla}{\vec{b}}$ 
involving the non-uniform MRI-supporting axisymmetric azimuthal 
field $\Boy\propto \cos{(k_{z_0} z)}$ characterized by $(m=0, n=0)$.
These couplings mediate the MRI. In all cases considered, we found 
that velocity perturbations are dominated by $(m=1,n=1)$, and magnetic 
perturbations by $(m=1,n=0)$ in the linear MRI growth phase.}}.
As they swing and their amplitude grows, these shearing wave packets
transfer energy back to $\Bo$ through nonlinear energy transfers in
the form of a quadratic axisymmetric electromotive force (EMF)
$\mathbf{\mathcal{\overline{E}}}_0=\overline{\mathbf{u}\times \mathbf{B}}$, 
sustaining it against resistive effect \textcolor{black}{(see Figs. 2 and 4
of H11 for detailed schematics and vizualisations of the self-sustained dynamics).}

The detailed dynamical complexity of $SN_{1m}$ and $SN_{2m}$ is significantly 
larger than that of cycles previously found in longer aspect ratio boxes, though:
\begin{enumerate}[i)]
\item the successive reversals of $\Bo$ are asymmetric in time.
\item each large-scale field reversal results from the accumulated
  \textcolor{black}{nonlinear self-interactions} of several successive 
  \textcolor{black}{MRI-unstable shearing wave packets}, regularly
  separated in time by $L_y/L_x S^{-1}$ (the quantized time
  separation is a consequence of the symmetries of the shearing box
  model). 
\item \textcolor{black}{Besides, different shearing wave packets have 
  a different energy content and polarization}, and therefore they do not 
  contribute equally to the large-scale field reversals.
\end{enumerate}

These different properties are illustrated in Fig.~\ref{fig_cycle_multish}.
In the case of $LB_{1m}$, the first reversal ($\Boy$ going from
positive to negative) is caused by the combined action of three
successive non-axisymmetric wave packets with different relative
amplitudes (represented by dark blue, blue and cyan curves in the bottom
panel). The second reversal also results from the combined action of
three distinct wave packets (represented by green to yellow colors)
with different relative amplitudes. The amplitudes of the second
series of waves are also different from those of the first series,
resulting in asymmetric $\Bo$ reversals.
For $LB_{2m}$, the asymmetry is even more pronounced: the first
reversal only involves two wave packets, while the second
involves three of them. Note that the period of $LB_{1m}$
and $LB_{2m}$ can be inferred directly from the product $L_y/L_x
S^{-1}$ times the number of waves \textcolor{black}{packets} 
implicated in the cycle, because
of the time quantization imposed by the symetries of the shearing box.
Finally, although the dynamics of the MRI-supporting field $\Bo$ is
governed by the cumulative action of shearing wave packets,
Fig.~\ref{fig_cycle_multish} shows that only a fraction of them are
significantly amplified by the MRI and contribute to the EMF. We find
that the evolution of these shearing wave packets, from the leading to 
the trailing phase, is influenced by nonlinear effects on
top of their linear MRI amplification.

\begin{figure*}
\centering
\includegraphics[width=0.94\textwidth]{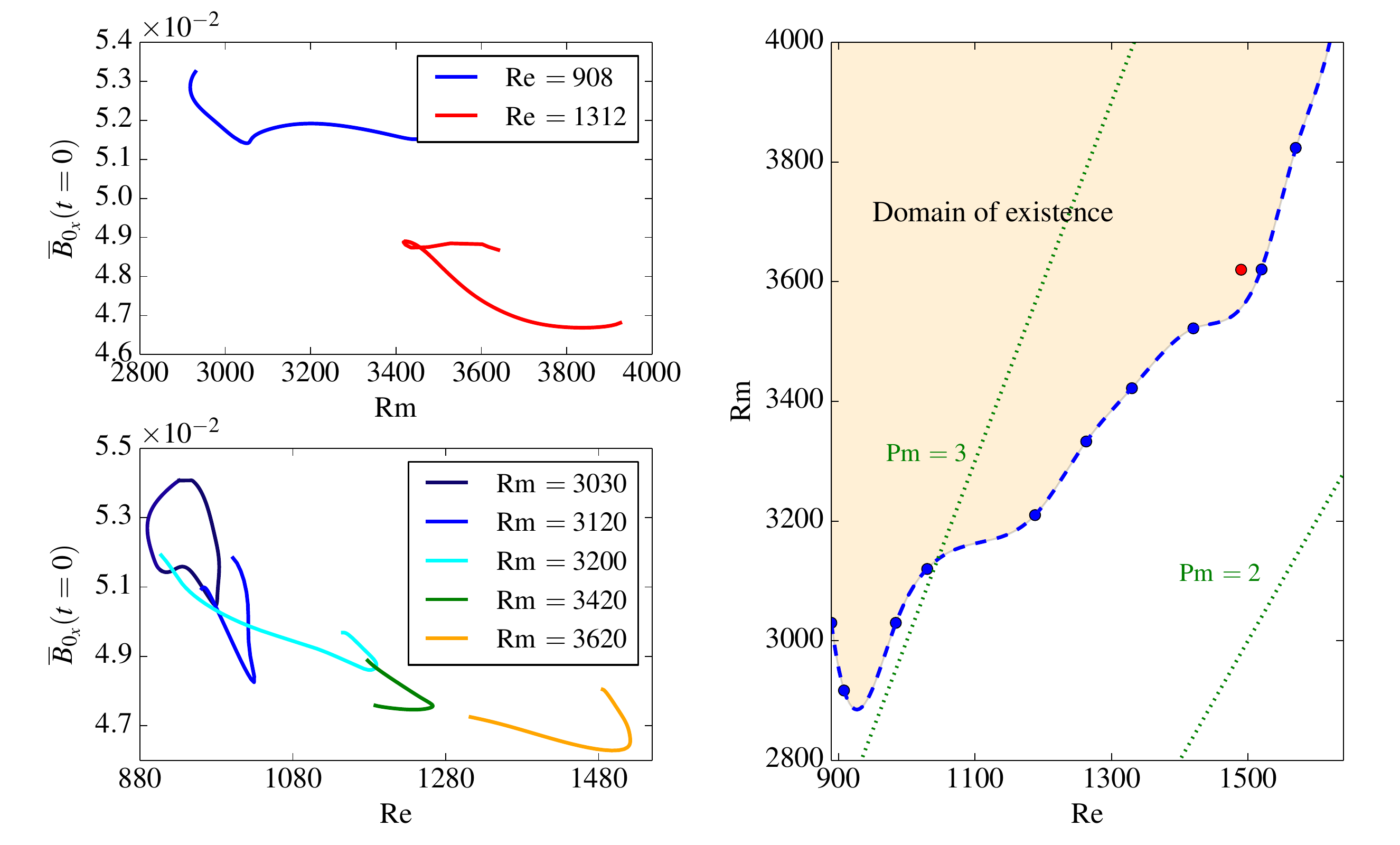}
 \caption{Left, top panel: selected continuation curves in Rm of
   $SN_{2m}$ at fixed Re; bottom panel: selected continuation curves
   in Re at fixed Rm. Right: existence boundary of $SN_{2m}$ in the
   (Re,Rm) plane (blue/dashed line) obtained {by interpolating} the different saddle
   node points $\text{Rm}_c(\text{Re})$ and
   $\text{Re}_c(\text{Rm})$ (blue bullets). The orange area shows the domain of
   existence of $SN_{2m}$ in parameter space. The red bullet shows a
   saddle note bifurcation point computed at double resolution
   ($96\times 96\times 144)$.}
\label{fig_cont_multish}
 \end{figure*}

The complex multiscale behaviour of these ``chimera'' cycles is
remarkable in the sense that it can still be understood in terms of
simple building blocks of a nonlinear self-sustanining MRI dynamo 
process, but is also reminiscent of the fully turbulent  butterfly
dynamo states observed in generic MRI dynamo simulations (see for
example the results of the incompressible simulations of
\citet{lesur08b} for $(L_x,L_y,L_z)=(0.5,2,1)$).  We will
come back to this at length in Sect.~\ref{dynamostat}.

The structure of these solutions finally suggests that self-sustained
dynamo cycles with an arbitrarily long period can be constructed from
successive MRI-unstable shearing wave packets (each new MRI-unstable
shearing wave packet involved in the dynamics of the chimera cycles is
generated consistently through a nonlinear physical process of
scattering of its predecessor off the radially modulated axisymmetric
magnetic field, see H11).

\subsection{Cycle continuations and existence boundaries}
In order to understand the physical origin of the dependence of the
MRI dynamo on dissipative processes, R13 and R15 performed
continuations in parameter space of different nonlinear 
cycles in elongated boxes, and found that their existence is
systematically limited to Pm larger than unity. Because of the
particular geometry used in these studies, it is natural to ask
whether these conclusions can be extended to moderate aspect ratio
boxes. We therefore performed a similar analysis for the more
complex $SN_{2m}$ pair in the
$(L_x,L_y,L_z)=(0.5,2,1)$ box, as this cycle exhibits all the
fundamental ingredients of the MRI dynamo process, and offers a good
example of self-sustained semi-statistical MRI dynamo behaviour in
moderate aspect ratio boxes. Although the full dynamics in this precise
geometry almost certainly involves many other cycles, this is the only
nonlinear periodic solution that we managed to compute and follow
accurately in parameter space for this box.

Figure \ref{fig_cont_multish} (left) shows selected
continuation curves for $SN_{2m}$. At fixed Re, these curves present a
turning point at a critical $\text{Rm}_c(\text{Re})$, confirming that
$SN_{2m}$ is born
out of a saddle node bifurcation. At fixed Rm, we find that this pair
of cycles exists only for a finite range of Re, whose upper bound
increases as Rm increases. The domain of existence of
$SN_{2m}$ in parameter space, shown in  Fig.~\ref{fig_cont_multish}
(right), is obtained by combining all the computed critical
$\text{Rm}_c(\text{Re})$ and $\text{Re}_c(\text{Rm})$.
The cycle existence is restricted to a region of Pm larger than unity
for the range of Rm that could be probed\footnote{These continuations
require a large computational effort. Each continuation curve in
Fig.~\ref{fig_cont_multish} (left) represents approximatively 10~000
direct numerical simulations at resolution $48\times48\times72$. In
addition, each branch can encounter many secondary bifurcations,
which makes it hard to follow and complete them. We
checked the numerical convergence of the results by
doubling the resolution in a few cases.}. Hence, the results
suggest that the conclusion of R13 and R15, that MRI dynamo
cycles are only sustained in the region of Pm larger than unity,
also hold in moderate aspect ratio boxes.

\section{Energy budgets and turbulent dissipation}
\label{budget}
Studying the magnetic energy transfers between different modes
involved in periodic nonlinear dynamics has previously proven useful
to identify possible physical reasons for the seeming decay of the MRI
dynamo at low Pm, and represents an important step towards a
rigorous statistical description of this kind of dyto be subjectnamo.
R15 showed that the subcritical "turbulent" magnetic diffusion
associated with the development of a nonlinear
cascade of MRI-unstable fluctuations to smaller scales at increasing
Re has a destructive effect on both the large-scale MRI-supporting
field and MRI-unstable fluctuations, and may explain the
disappearance of dynamo cycles at moderate Rm and low Pm in large
aspect ratio boxes. Here, we aim to determine if the same conclusions
hold for moderate aspect ratio dynamics by examining the
{energy budget} of the saddle node pair $SN_{2m}$. This analysis is
rather technical, and may be skipped by readers not interested in
details. The conclusions are qualitatively the same as in 
R15, and are summarized in Sect.~\ref{conclusions}.

\subsection{Energetics of the MRI supporting field $\Bo$}
\label{Bobudget}
\begin{figure}
\centering
\includegraphics[width=\columnwidth]{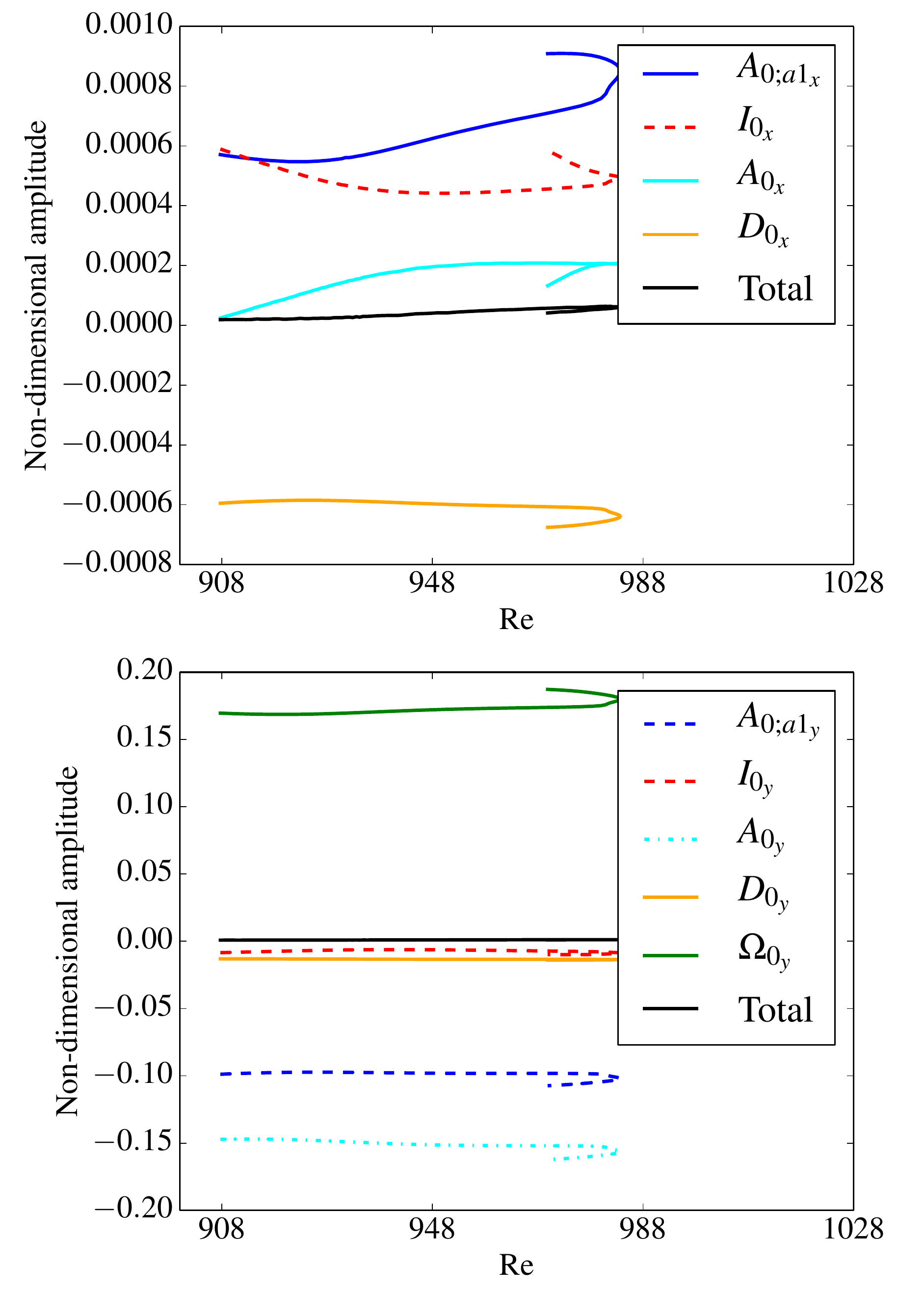}
 \caption{$x-$ (top) and $y-$ (bottom) projections of the energy
   budget~(\ref{B0budget}) of the axisymmetric field component
   $\Bo$ of $LB_{2m}$ as a function of Re ($\text{Rm}=3030$).}
\label{fig_budget_B0}
 \end{figure}

The MRI dynamo intrinsically relies on the sustainment of a
large-scale MRI-supporting  magnetic field component against
dissipative processes, which in shearing box simulations is 
dominated by the axisymmetric $\Bo$. It is then of primary
importance to analyse the energy budget of $\Bo$ over a full 
cycle period $T_{SN_{2m}}=5\,S^{-1} L_y/L_x=20\, S^{-1}$,
\begin{equation}
\label{B0budget}
{\mathbf{\boldsymbol{\Omega}}}_0+{\mathbf{I}}_0+{\mathbf{A}}_0+{\mathbf{D}}_0=\bf{0}~,
\end{equation}
where
\begin{equation}
\mathcal{\mathbf{\boldsymbol\Omega}}_0=
-S\langle\overline{B}_{0_y}\overline{B}_{0_x}\rangle\mathbf{e}_y,
\quad  \quad
\mathcal{\mathbf{I}}_0= \langle \Bo  \circ \overline{\mathbf{B}\cdot \nabla\mathbf{u}} \rangle,
\end{equation}
\begin{equation}
 \mathcal{\mathbf{D}}_0=-\left({k_{z_0}^2}/{\Reym}\right) \langle  \Bo \circ \Bo  \rangle,  \quad  \quad
 \mathcal{\mathbf{A}}_0=
-\langle \Bo \circ  \overline{\mathbf{u}\cdot \nabla\mathbf{b}} \rangle
 \end{equation}
and $\langle\rangle$ denotes the volume and cycle period average 
and $\circ$ the entrywise
product \textcolor{black}{[$(\mathbf{X}\circ \mathbf{Y})_i=X_i Y_i$]}. $\mathbf{\boldsymbol{\Omega}}_{0}$, $\mathbf{I}_{0}$,
$\mathbf{A}_{0}$ and $\mathbf{D}_{0}$  are the energetic contributions
to $\Bo$ respectively associated to the $\Omega$-effect, nonlinear 
induction, nonlinear advection and ohmic dissipation. We also define
${\mathbf{A}}_{0;m_i}$, the magnetic energy exchanged through 
nonlinear advection between $\Bo$ and a given non-axisymmetric 
\textcolor{black}{shearing wave packet}, labelled $m_i$. 
Figure~\ref{fig_budget_B0} shows
the $x$ and $y$ projections of Eq.~(\ref{B0budget}) for the lower
branch of $SN_{2m}$ as a function of $\text{Re}$, at fixed
$\text{Rm}=3030$ (the upper
branch behaves similarly). The MRI-supporting azimuthal field
$\Boy$ loses most of its energy through the nonlinear
advective transfer ${A}_{0_y}<0$, which acts as a weakly nonlinear
``turbulent'' diffusion. The laminar ohmic dissipation ${D}_{0_y}$ is
negligible. The $\Omega$ effect is the only net source term for
$\Boy$, indicating that the sustainment of the radial $\Bor$
is critical for the dynamo process as a whole. 
Figure~\ref{fig_budget_B0} (top) shows that $\Bor$ gains
energy from both nonlinear induction $I_{0_x}$ and advection $A_{0_x}$,
and dissipates a large amount of this energy directly through ohmic
diffusion. The part of ${A}_{0x}$ associated with the nonlinear
correlations of the \textcolor{black}{dominant MRI-unstable $(m=1,n=1)$ 
velocity perturbation and $(m=1,n=0)$ magnetic perturbation}
(denoted by ${A}_{{0;a1}_x}>0$ in the figure, anticipating the notations
introduced in Sect.~\ref{formalism} below) is positive, and much
larger than the total $A_{0_x}$. This implies that some energy is also
transferred nonlinearly from $\Bor$ to other smaller-scale modes, which
can be interpreted again as a turbulent diffusion acting on $\Bor$.
 
This budget is reminiscent of the results of R15 for large aspect
ratio cycles. The most notable difference is that $I_{0_x}$ is not
small here, and is of the order $A_{0_x}$. \textcolor{black}{While 
this term may also be important for the sustainment of 
the dynamo and for the Pm problem (as Re increases, $I_{0_x}$ 
decreases while $A_{0_x}$ increases), in the following we 
focus on the role of $A_{0_x}$ in relation to the results of R15.} 

\subsection{Energetics of shearing wave packets}

\begin{figure}
\centering
\includegraphics[width=\columnwidth]{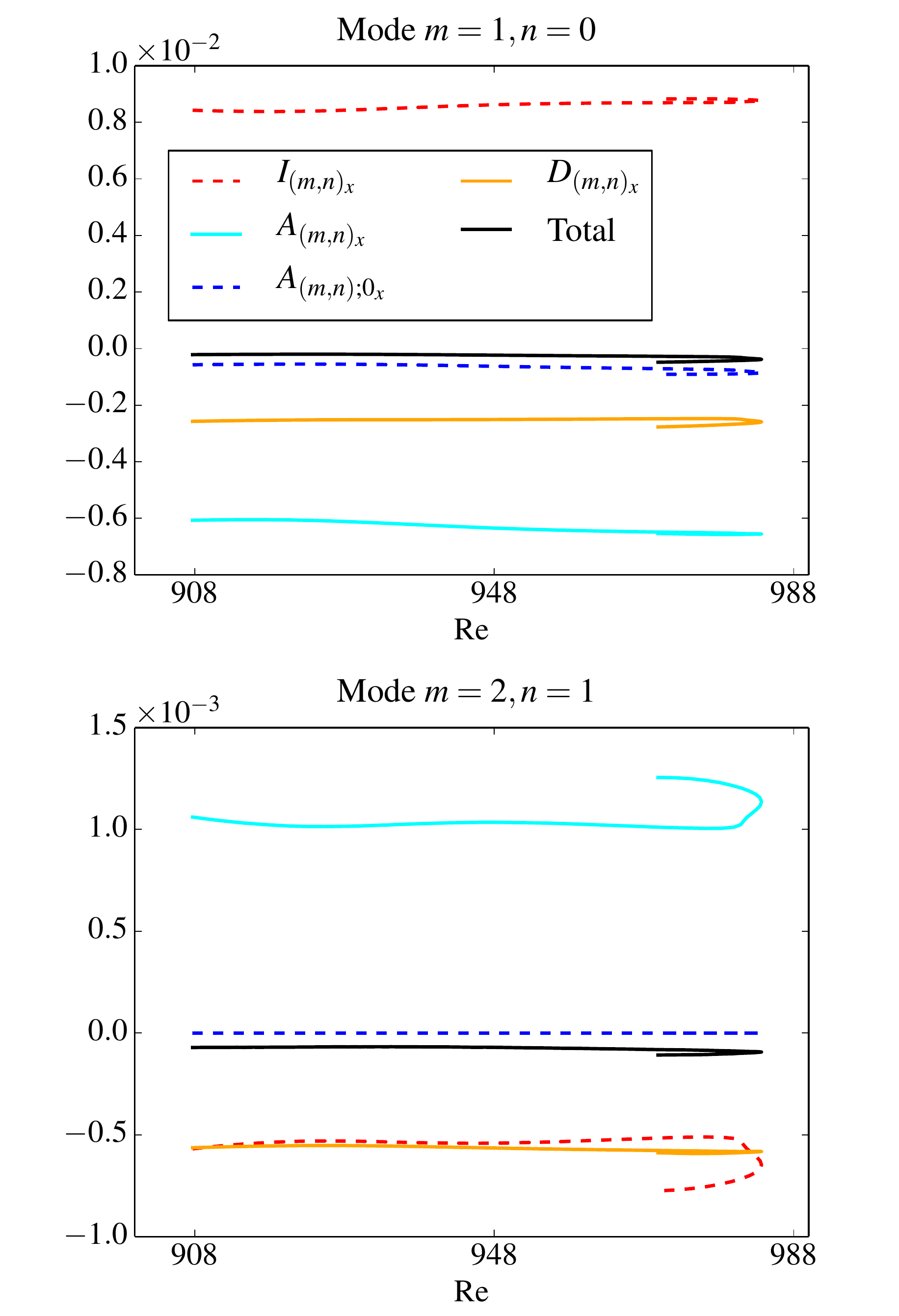}
 \caption{$x$-projection of the cumulated magnetic energy
   budgets~(\ref{bnonaxibudget}) of non-axisymmetric perturbations
   as a function of Re, for the lower branch $LB_{2m}$
   at $\text{Rm}=3030$. Top: active MRI-unstable \textcolor{black}{perturbation} 
   $a_1$ \textcolor{black}{(magnetic perturbation $m=1$, $n=0$)}. Bottom: 
   slaved MRI-stable \textcolor{black}{magnetic perturbation} ($m=2$, $n=1$). 
   The sum of all source and dissipative terms remains close to 0.}
\label{fig_budget_as}
 \end{figure}

In the previous analysis, the key quantity $A_{0_x}$ included both the
(positive) energy exchanged nonlinearly between $\Bor$ and MRI-unstable
\textcolor{black}{perturbations} and the (negative) energy exchanged 
between $\Bor$ and other
smaller-scale \textcolor{black}{perturbations}, integrated over a full 
period of $SN_{2m}$. To
understand the details of these energy transfers in moderate aspect
ratio boxes, we now investigate the magnetic energy budget of
individual non-axisymmetric wave packets ($m\neq 0$).
Wave packets with different lagrangian radial wavenumbers
$\ell$ differing by $\Delta \ell=1$ swing in the box one after the other
at regularly spaced time intervals, and we label each successive
packet by its superscript $\ell$. For $SN_{2m}$, $\ell$ varies
from 0 to 4 (the evolution of total energy of each of these waves 
is shown in Fig.~\ref{fig_cycle_multish}). $m=1$ wave \textcolor{black}{packets}, 
which dominate the dynamics here, have a typical swinging time $S^{-1} L_y/L_x$. 
The magnetic energy budget \textcolor{black}{for a given $(m,n)$}, 
summed over all successive $\ell$ waves, reads 
\begin{align}
\label{bnonaxibudget}
\mathbf{\boldsymbol{\Omega}}_{(m,n)}+\mathbf{I}_{(m,n)}+\mathbf{A}_{(m,n)}+\mathbf{D}_{(m,n)}&
\simeq  \sum_\ell
\left<\mathbf{b}^\ell_{(m,n)}\circ\frac{\partial}{\partial t}
    \mathbf{b}^\ell_{(m,n)}\right> \nonumber\\
 &\simeq \bf{0}.
\end{align}
Here, $\mathbf{\boldsymbol{\Omega}}_{(m,n)}$, $\mathbf{I}_{(m,n)}$,
$\mathbf{A}_{(m,n)}$ and $\mathbf{D}_{(m,n)}$ are energetic
contributions associated to the $\Omega$-effect, nonlinear 
induction, nonlinear advection and ohmic dissipation,
\begin{equation}
\mathbf{\boldsymbol{\Omega}}_{(m,n)}=-\sum_\ell
S\langle\overline{b}^\ell_{{(m,n)}_x}\overline{b}^\ell_{{(m,n)}_y}\rangle\,\mathbf{e}_y,
\end{equation}
\begin{equation}
\mathbf{I}_{(m,n)}= \sum_\ell\langle
\mathbf{b}^\ell_{(m,n)}\circ(\mathbf{B}\cdot\nabla
\mathbf{u})^\ell_{(m,n)}\rangle,
\end{equation}
\begin{equation}
\mathbf{D}_{(m,n)}= -\sum_\ell \eta k^2\,\langle \mathbf{b}^\ell_{(m,n)}
\circ \mathbf{b}^\ell_{(m,n)}\rangle,
\end{equation}
\begin{equation}
\mathbf{A}_{(m,n)}= \sum_\ell\langle
\mathbf{b}^\ell_{(m,n)}\circ(\mathbf{u}\cdot\nabla
\mathbf{B})^\ell_{(m,n)}\rangle.
 \end{equation}
 The advective term $\mathbf{A}_{(m,n)}$ basically represents the magnetic
 energy exchanged between $(m,n)$ and all the other modes. The budget
 is only approximatively zero because each $\ell$ waves carries a tiny
 amount of energy in the strongly leading and trailing phases, and has
 (very weak) nonlinearly couplings to its predecessors and successors
 (see H11). These couplings can be neglected in the context of this
 particular energetic analysis. We now analyse the energy budget of
 two illustrative ``active'' and ``slaved'' non-axisymmetric 
 \textcolor{black}{perturbations} taking part in the dynamics of $LB_{2m}$. 

\subsubsection{Active \textcolor{black}{perturbations}}
We first consider \textcolor{black}{magnetic 
perturbations} with ($m=1$, $n=0$).
\textcolor{black}{As explained in the footnote in Sect.~\ref{cycles}, 
these perturbations, in conjunction with ($m=1$, $n=1$)
velocity perturbations, dominate the non-axisymmetric
MRI-unstable dynamics in the system under consideration.
They are subject to energy injection through
induction and, through their non-linear EMF coupling
$\overline{\vec{u}\times\vec{b}}$ to $(m=1,n=1)$ 
velocity perturbations, act as the main energy 
provider of $\Bo$ through nonlinear transfers.
We therefore label this set of non-axisymmetric 
perturbations as $a_1$ for ``active''.}

\textcolor{black}{The energetics of the radial component
of these magnetic perturbations}
is illustrated in Figure~\ref{fig_budget_as} (top), which shows the 
radial projection of Eq.~\eqref{bnonaxibudget} for $(m=1,n=0)$ a 
function of Re. As expected of an MRI-unstable situation, energy 
is gained through the induction term $(I_{{a_1}_x}>0)$, and 
redistributed through nonlinear advective transfers $(A_{{a_1}_x}<0)$ 
to both $\Bo$ and smaller-scale modes. \textcolor{black}{The azimuthal 
component (not shown) extracts energy through the $\Omega$-effect.}
Other $m=1$ \textcolor{black}{magnetic modes} with $n=2,4$ behave in the 
same way. 

A short digression is in order. Here and before, we have taken a
Eulerian viewpoint, which is easier to analyse numerically but not
necessarily  very transparent physically. The previous
results simply mean that the large-scale ``lagrangian'' magnetic field,
described in a good first approximation by the sum of $\Bo$ and
shearing non-axisymmetric $m=1$ waves (and multiple $n$)
at the Re and Rm considered, is basically sustained by MRI induction
($I_{a_1x}$). The nonlinear advective transfers between $\Bo$ and
$a_1$ redistribute energy between these two Eulerian modes but
conserve their total energy  (see H11, Fig. 4 for a lagrangian picture
of MRI dynamo reversals).

\subsubsection{Slaved \textcolor{black}{perturbations}}
The previous analysis explains how the large-scale field can be 
sustained through a dynamo process \textcolor{black}{involving 
non-axisymmetric MRI-unstable wave packets}. \textcolor{black}{However, 
not all perturbations that take part in the nonlinear dynamics of $LB_{2m}$
are excited by the MRI nor feed back energy into the large-scale 
dynamo field. As can be seen in Fig.~\ref{fig_budget_as} (bottom), 
magnetic perturbations such as $m=2$, $n=1$ 
have $I_{(m,n)_x}<0$ and $A_{(m,n)_x}>0$ (a similar budget
is obtained for $m\geq 2$ and $n\geq 1$). Such "slaved" perturbations
represent smaller-scale structures which are not themselves amplified 
by the MRI.} They only gain energy through nonlinear transfers from 
larger-scale MRI-amplified active \textcolor{black}{perturbations}, and 
dissipate it through laminar dissipation. In other words, they act 
as a ``turbulent'' diffusion for the large scale field and are therefore 
destructive for the dynamo. We will use this separation between ``active'' 
and ``slaved'' \textcolor{black}{perturbations} in the next paragraph to 
quantify this diffusion in a global way as function of Re and Rm.

\begin{figure*}
\centering
\includegraphics[width=1.\textwidth]{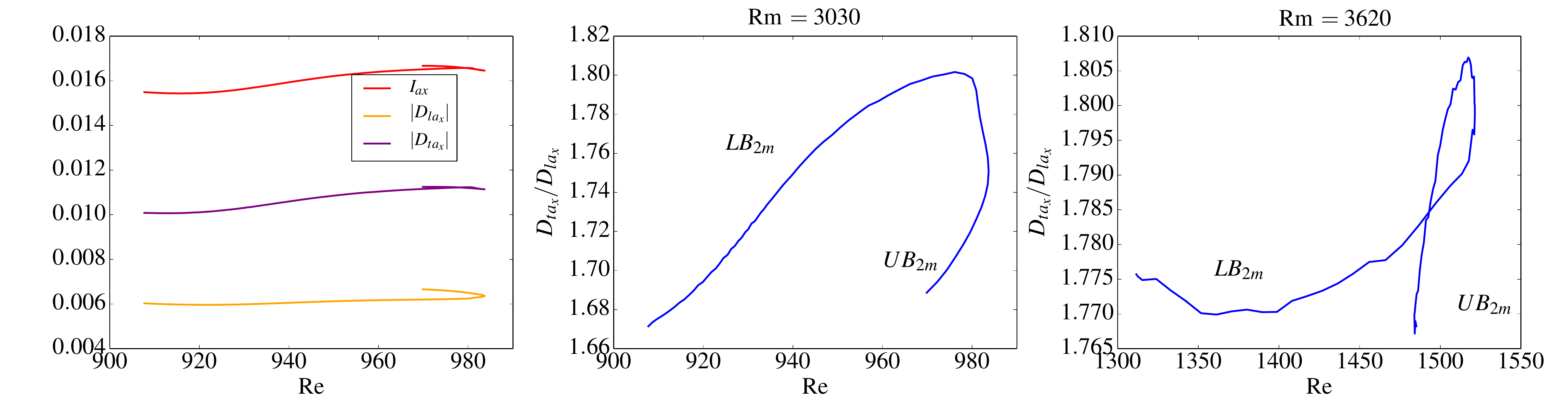}
\caption{Left panel: the different terms $I_{a_x}$, $D_{la_x}$ and $D_{ta_x}$ in
the net magnetic energy budget~(\ref{bilan_glob_multish}) for
$LB_{2m}$ as a function of Re, at fixed $\text{Rm}=3030$. Middle and
right panels: ratio $D_{ta_x}/D_{la_x}$ for the lower and upper branches
of $SN_{2m}$ as a function of Re, for $\text{Rm}=3030$ and
$\text{Rm}=3620$ respectively (the roughness of the rightmost 
plot is due to the limited time-sampling of simulation outputs used 
in the postprocessing analysing stages, not to the numerical resolution
of the simulations).}
\label{fig_budget_glob1}
 \end{figure*}

\subsection{Characterization of turbulent magnetic dissipation}
\subsubsection{Formalism}
\label{formalism}
We now use the previous concept of active and slaved 
\textcolor{black}{perturbations}
to characterize turbulent magnetic diffusion more quantitatively,
and to study its impact on the dynamo at large Re (low Pm). 
A Fourier mode will be considered as active if
\begin{equation}
I_{{(m,n)}_x}>0  \\
 \text{and}\\
A_{(m,n)_x}<0~.
\end{equation}
These modes will be labelled $a_i=1,..., N_a$. On the
contrary, a \textcolor{black}{magnetic} mode will be considered as slaved if
$I_{(m,n)_x}<0$ or $A_{(m,n)_x}>0$. Slaved modes are stabilized with
respect to the MRI by a stronger diffusion and magnetic tension,
and will be labelled  $s_i=1,..., N_s$.  ``Small-scale'' axisymmetric
modes ($m=0$, $n>1$) will also be considered as slaved because they
do not contribute directly to the sustainment of $\Bo$.

We can then think of turbulent diffusion as the cumulated nonlinear
effect of all slaved modes on all the active ones and on $\Bo$. 
The same idea was used in R15 to characterize turbulent diffusion for
large aspect ratio cycles. The only difference here is that the
physical structure of non-axisymmetric active MRI \textcolor{black}{perturbations} 
taking part in the dynamics of chimera cycles is more complex (it
involves a larger number of $m$ and $n$ Fourier modes), and that
these cycles involve a succession of such non-axisymmetric structures
(different $\ell$, see Fig.~\ref{fig_cycle_multish}).

We define the amount of energy exchanged during a cycle period
between a mode $(m,n)$ and a mode $(m',n')$:
\begin{equation}
\mathbf{A}_{(m,n);(m',n')}=-\langle
\mathbf{b}_{(m,n)} \circ  ({\mathbf{u}_{(m-m',n-n')}\cdot \mathbf{\nabla}\,\mathbf{b}_{(m',n')})} \rangle
\end{equation}
with
\begin{equation}
\label{eq_revers}
\mathbf{A}_{(m,n);(m',n')}= -\mathbf{A}_{(m',n');(m,n)}~.
\end{equation}
We then have the following relation, using the active/slaved shorthand notation:
\begin{equation}
\label{advectionB0}
{\mathbf{A}}_0=\sum_{j=1}^{N_a}{\mathbf{A}}_{0;a_j}+\sum_{j=1}^{N_s}{\mathbf{A}}_{0;s_j}~,
\end{equation}
\begin{equation}
\label{advection_ai}
\mathbf{A}_{a_i}=\mathbf{A}_{a_i;0}+\sum_{j\neq i} \mathbf{A}_{a_i;a_j}+\sum_{j=1}^{N_s}\mathbf{A}_{a_i;s_j}~,
\end{equation}
which simply expresses that any active mode exchanges energy with the
background field $\Bo$, the other active modes and the slaved modes. 
We define the turbulent dissipation $\mathbf{D}_{ta}$ as the total magnetic
energy nonlinearly transferred from the system \{$\Bo$+active modes\}
to the slaved modes: 
\begin{equation}
\label{Dtequation}
\mathbf{D}_{ta}=\sum_{j=1}^{N_s}\mathbf{A}_{0;s_j}+\sum_{i=1}^{N_a} \sum_{j=1}^{N_s}\mathbf{A}_{a_i;s_j}~.
\end{equation}
We now focus on the $x$-component of Eq.~(\ref{Dtequation}), as it 
is the most critical for the sustainement of the dynamo (the
$y$-component of the magnetic field always benefits from the
$\Omega$-effect). Summing Eq.~(\ref{advection_ai}) over all the
active modes, using Eq.~(\ref{advectionB0}) and the symmetry 
condition (\ref{eq_revers}), we obtain
\begin{equation}
D_{ta_x}=A_{0_x}+\sum_{j=1}^{N_a} {A}_{{a_j}_x}~.
\end{equation}
Finally, summing the energy budget~(\ref{bnonaxibudget}) over all
the active modes, and adding the contribution of the
axisymmetric field $\Bo$ to the result, we obtain the 
effective magnetic energy budget of the active radial 
field component,
\begin{equation}
\label{bilan_glob_multish}
\underbrace{\left(I_{0_x}+\sum_{i=1}^{N_a} {I}_{{a_i}_x}\right)}_{I_{a_x}}\,+\,\underbrace{\left(D_{0_x}+\sum_{i=1}^{N_a} {D}_{{a_i}_x}\right)}_{D_{la_x}}\,+\,D_{ta_x} \simeq  0~.
\end{equation}
The first curly brace term $I_{a_x}$ is the net magnetic energy
injected in the system \{$\Bo$+active modes\} over a cycle period. The
second curly brace term $D_{la_x}$, is the magnetic energy directly 
lost by $\Bo$ and active modes through ``laminar'' ohmic dissipation.
Finally, by construction and as desired, $D_{ta_x}$ stands for the
energy lost by these modes through nonlinear energy
transfers to smaller scales. This equation \textcolor{black}{indicates} that the
magnetic energy injected via the MRI into the active dynamo field
balances the sum of laminar and turbulent magnetic dissipations 
(for a periodic solution). 

\subsubsection{Turbulent magnetic dissipation for $SN_{2m}$}
Having been through great pains to formalise the analysis of the magnetic
energy budgets of the MRI dynamo in the shearing box, we can now apply
it to the $SN_{2m}$ chimera cycles. Figure \ref{fig_budget_glob1}
(left) shows the induction term $I_{a_x}$ and the two dissipative terms
$D_{la_x}$ and $D_{ta_x}$ for this pair of cycles as a function of Re, at fixed
$\text{Rm}=3030$. It appears that the turbulent dissipation $D_{ta_x}$ 
for both lower and upper branches increases significantly more rapidly
than the laminar dissipation $D_{la_x}$ as Re is increased. As shown in
Fig.~\ref{fig_budget_glob1} (middle and right panels), this ratio
$D_{ta_x}/D_{la_x}$ increases by 7\% from $\text{Re}=910$ to
$\text{Re}=980$. The same result holds at the larger $\text{Rm}=3620$
at which the cycle is sustained for a
wider range of Re. These results therefore suggest that the
disappearance of cycles at large Re is caused by the relative
enhancement of turbulent magnetic dissipation associated with the
development of smaller-scale velocity fluctuations. At larger Re,
this additional dissipation can only be compensated if we can make 
induction larger relative to laminar dissipation, i.e. by going to
larger Rm. This seems to explain well the typical existence boundary
of MRI dynamo cycles \textcolor{black}{in the (Re,Rm) plane}.

\subsubsection{Illustration of turbulent magnetic dissipation in
  aperiodic test simulations}
\begin{figure}
\centering
\includegraphics[width=\columnwidth]{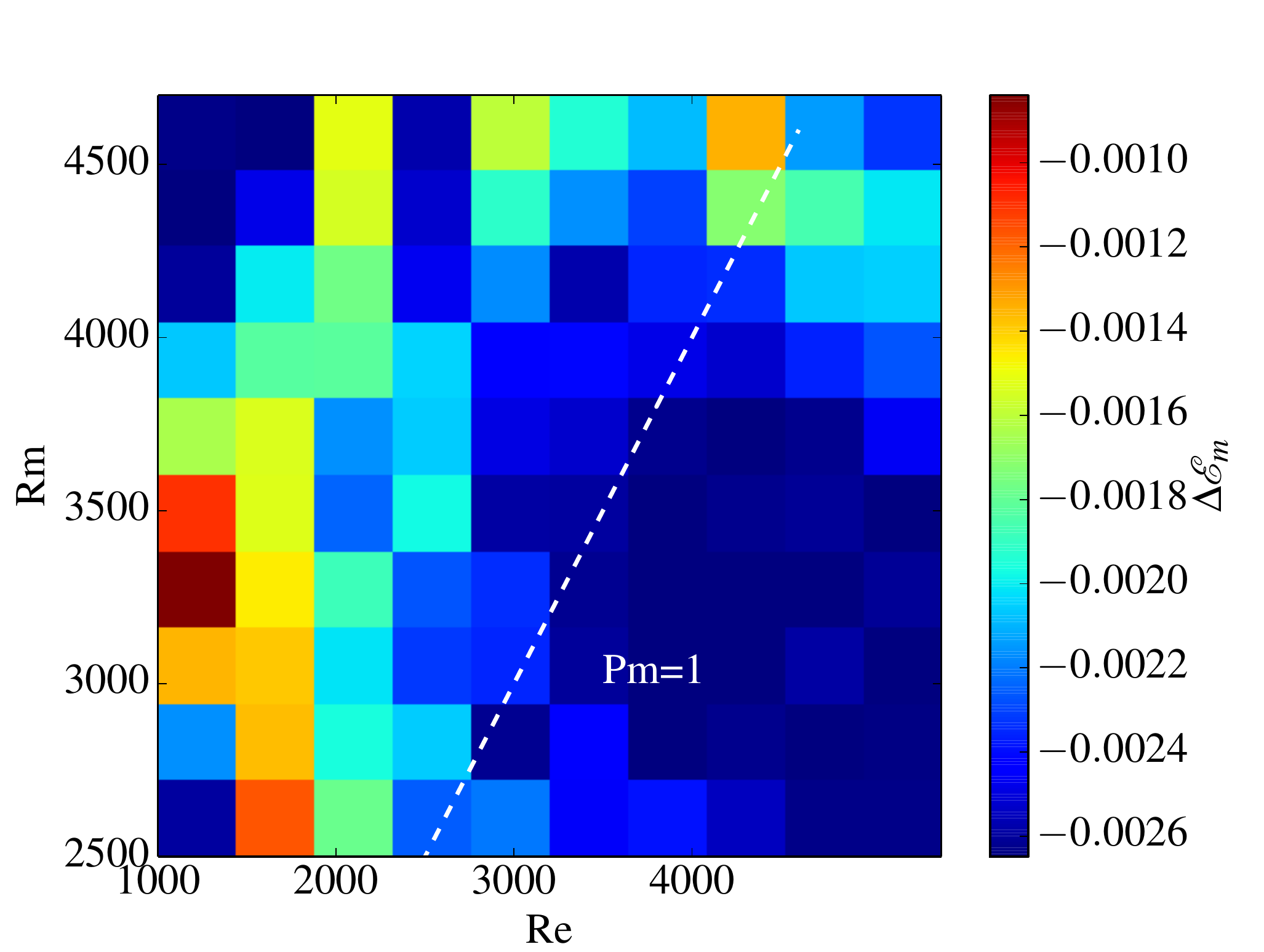}
 \caption{Magnetic energy change $\Delta \mathcal{E}_m$ of the
   axisymmetric radial magnetic field component $\Bor$ between $t=0$
   and $T_{SN_{2m}}=5 S^{-1} L_y/L_x$ for simulations at different Re
   and Rm initialized with the same initial condition consisting of
   the MHD state $LB_{2m}(t=0)$ at $\text{Re}=908$ and
   $\text{Rm}=3030$.}
\label{fig_Dem}
\end{figure}

To investigate whether the conclusions of the previous analysis
pertain to more general circumstances and can explain the seeming
disappearance of the MRI dynamo as a whole at low Pm, we
finally considered slightly more generic aperiodic test
simulations in the phase-space vicinity of $SN_{2m}$. 
We performed a series of simulations at different Re and Rm
initialized with the same initial condition, consisting of the initial
state  of the $LB_{2m}$ cycle computed at $\text{Re}=908$ and
$\text{Rm}=3030$. Except for these precise values
of Re and Rm, the time evolution from this initial condition is 
not periodic and can result in either a gain or loss of magnetic
energy after an integration of the equations over a period of the
original cycle $T_{SN_{2m}}=5S^{-1} L_y/L_x$. We then computed the
magnetic energy change $\Delta \mathcal{E}_m$ in the $\Bor$ component
after this time, as a function of Re and Rm. Figure
\ref{fig_Dem} shows that $\Delta \mathcal{E}_m$ is always negative
for $\text{Re}>908$, and that the energy loss increases with Re at
fixed Rm (or equivalently lower Pm). Moreover, the nonlinear
term $A_{0_x}$ responsible for the sustainement of $\Bor$ decreases
with Re. Although the latter result does not rigorously quantify
the  turbulent dissipation, it indicates that some magnetic energy
injected into  MRI-unstable waves is transferred to small scales
rather than into $\Bor$ at large Re, resuting in the decay of the
MRI-supporting field and therefore of the dynamics as a whole.

\begin{figure*}
\centering
\includegraphics[width=\textwidth]{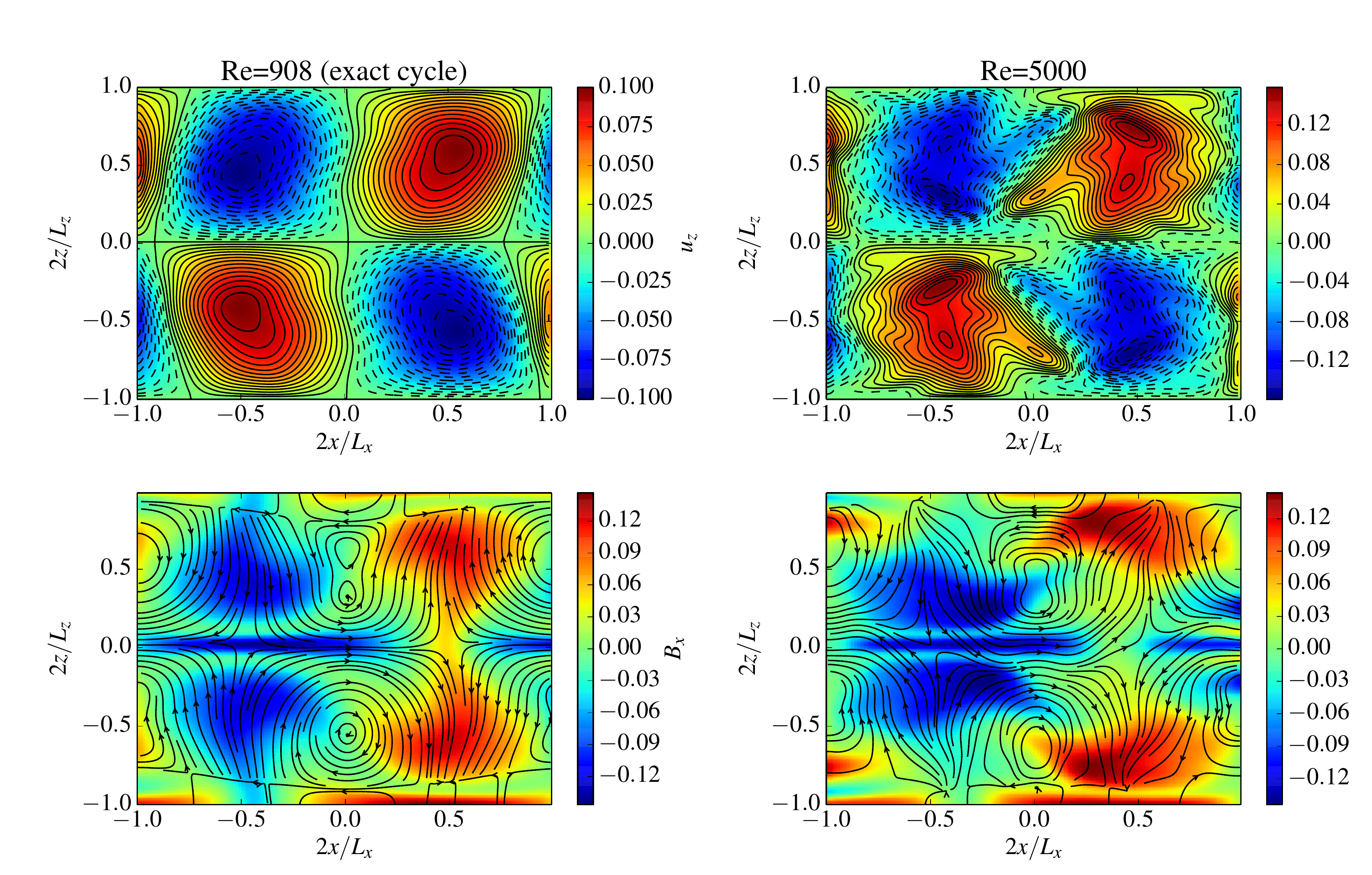}
 \caption{Color snapshots of the vertical velocity field $u_z$ (top)
   and radial magnetic field $B_x$ (bottom) in the poloidal plane
   ($x$,$z$). Left panel: $LB_{2m}$ at $\text{Re}=908$. Right panel: 
   MHD state integrated from the same initial condition, but at
   $\text{Re}=5000$. $\text{Rm}=3030$ in both cases.
The snapshots are taken at the time $t=5\,S^{-1}$
   around which the first $\Bo$ reversal of $LB_{2m}$ occurs. The
   dashed black lines are iso-contours of $u_z$, the full black lines
   with arrows  are poloidal velocity streamlines.}
\label{fig_surface_flux}
 \end{figure*}

Figure~\ref{fig_surface_flux} provides a direct physical illustration of
the effects of turbulent magnetic diffusion at large Re. The
visualizations show the vertical velocity field and total radial
magnetic field $B_x$ in a poloidal $(x,z)$ plane around the time
of reversal of $\Bo$. At $\text{Re}=5000$,
vertical velocity field perturbations clearly have more small-scale
structure than at $\text{Re}=908$, and the non-axisymmetric counter-rotating 
flow vortices driving the reversal of the large-scale field (see H11) are
much less regular. The magnetic field inherits some of this small-scale
structure through ``turbulent'' advection. The magnetic gradients
(electric currents) are clearly larger at $\text{Re}=5000$ than at
$\text{Re}=908$, leading to enhanced magnetic dissipation.
\section{The missing link to turbulent accretion disk dynamo models ?\label{dynamostat}}

As shown in Sect.~\ref{cycles}, chimera dynamo cycles can be
understood in terms of simple physical building blocks of a nonlinear
self-sustanining MRI dynamo process, but are
also reminiscent of the statistical butterfly states observed in many
simulations. The structure of chimeras suggests that
cycles made of a large number of MRI-unstable shearing wave packets
(dynamo ``quanta'') can be constructed in shearing boxes of different
aspect ratio (at least for Pm>1).
Although they do not accomodate all the physical elements entering
the accretion disk dynamo problem (such as magnetic buoyancy), the
dynamical behaviour of these structures therefore raises the prospect
that a statistical theoretical MRI dynamo model can be rigorously
derived from first principle physics. The aim of this section is to
highlight some important  dynamical features of the solutions at hand,
some but not all of which are already taken into account  in existing
theoretical models
\citep{lesur08,lesur08b,gressel10,gressel15,squire15}, in order to
provide constructive guidance for future work on statistical theory.

\subsection{Statistical linearity vs dynamical nonlinearity}
\begin{figure}
\centering
\includegraphics[width=\columnwidth]{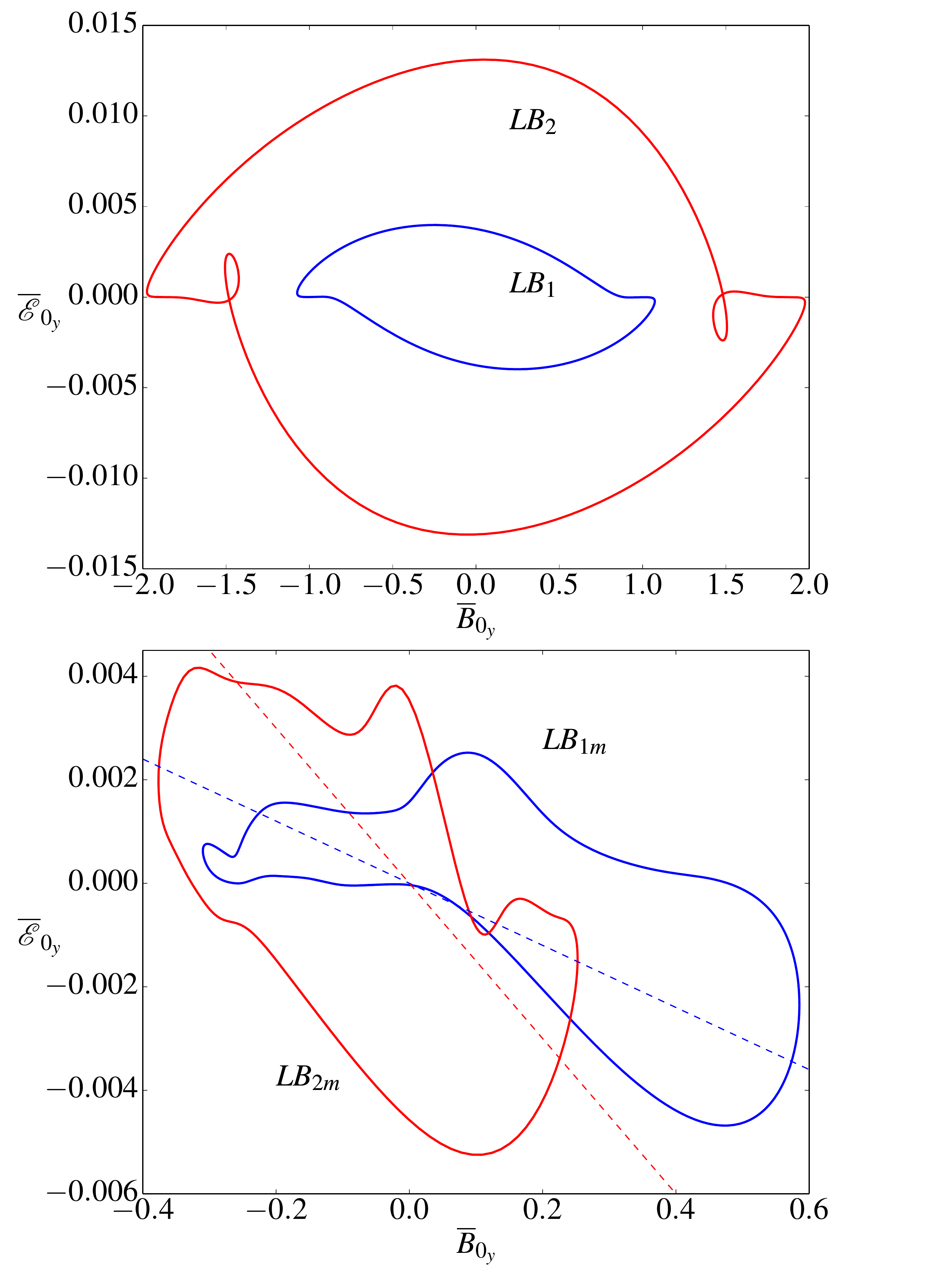}
\caption{Phase portraits of periodic dynamo solutions in the
  $\mathcal{E}_{0_y}$ vs $\Boy$ plane (large-scale axisymmetric EMF vs
  large-scale axisymmetric azimuthal field). Top: lower branches of
  the two cycles $LB_1$ and $LB_2$ computed in a large
  aspect ratio box $L_y/L_x=28.57$ (see H11 and R13).
  Bottom: new chimera lower branches $LB_{1m}$ and $LB_{2m}$ computed
  in moderate aspect ratio boxes $(0.7,6,2)$ and $(0.5,2,1)$. 
  \textcolor{black}{The dashed lines indicate the linear trend 
   between $\mathcal{E}_{0_y}$ and $\Boy$.}}
\label{fig_emf_multish}
\end{figure}

To introduce the matters, we reproduce in Fig.~\ref{fig_emf_multish}
(top) the instantaneous relationship between the large-scale
axisymmetric nonlinear EMF $\mathbf{\mathcal{\overline{E}}}_0$ and
$\Bo$ for the lower branch cycles $LB_1$ and $LB_2$ computed by H11
and R13 in azimuthally elongated shearing boxes. As noted in H11, the 
standard mean-field theory ansatz of a linear relationship
(\citealt{steenbeck66,moffatt77,branden05}, see
\citealt{gressel10,blackman12,gressel15} for applications to the
accretion disk dynamo context) doesn't fit these solutions of the full
MHD equations (a similar mismatch has been reported for a
magnetic-buoyancy driven dynamo, see \cite{cline03}). A possible
explanation for this discrepancy is that such solutions, despite
being nonlinear, are not turbulent or statistical in essence, and that
the linear mean-field relationship only holds statistically. The
question nevertheless arises as to whether and how one can rigorously
relate the fundamental linear
and nonlinear physical processes illustrated by these
three-dimensional solutions to a seemingly more abstract
statistical two-dimensional effective description. Of particular
concern with the mean-field description in the context of
instability-driven dynamos (such as the MRI dynamo) is the lack 
of explicit connection between mean-field effects
and fundamental dynamical processes (such as the MRI) in the 
absence of which there can be no dynamo-generating turbulence at all
in the first place in the corresponding systems. 

This possible limitation of classical mean-field theory has motivated
the development of alternative quasi-linear models of the turbulent
MRI dynamo describing the
cumulated effects of a statistical assembly of shearing waves whose
individual physical linear evolution can be computed consistently
either analytically or numerically \citep{lesur08b,lesur08,squire15}. 
The structure of our chimera dynamo cycles qualitatively vindicates 
this approach (albeit with a few caveats to be discussed below). 
Another encouraging sign that it may be possible to close in on
a fully self-consistent statistical theory is illustrated by
Figure~\ref{fig_emf_multish} (bottom), which now shows the
instantaneous relationship between the large-scale axisymmetric
nonlinear $\mathbf{\mathcal{\overline{E}}}_0$ and $\Bo$ for the
lower branch chimera cycles $LB_{1m}$ and $LB_{2m}$. While the 
detailed relationship remains nonlinear, a clear average linear 
trend  emerges in comparison to the $LB_1$ and $LB_2$ cases. This
result supports the common intuition that complex three-dimensional
nonlinear multiscale dynamics tend to generate statistically simple
effective large-scale dynamical states in the turbulent limit.
\subsection{Roles of active and passive \textcolor{black}{perturbations}}

The most direct explanation for a statistical correlation between 
the large-scale EMF and large-scale magnetic field in the MRI 
dynamo problem can be found in the quasi-linear theory of
\cite{lesur08b}, which predicts a linear dependence between the
MRI-supporting field and the nonlinear EMF generated by MRI-amplified
shearing waves packets. This is notably expected if the MRI-supporting
toroidal field $\Boy$ and azimuthal wavenumbers of the shearing waves
are such that the MRI is on its weak-field branch, so that the MRI
growth rate is proportional to $\Boy$. In a simulation, one would
therefore expect that successive shearing waves ``see'' a slowly
time-evolving large-scale field and that their relative growth
(and ensuing nonlinear EMF feedback) is modulated according 
to the amplitude of the MRI-supporting field at the time of their
amplification, resulting in the linear trend discussed above. This correlation is not observed for individual $\ell$ shearing wave
packets of our chimera cycles (individual wave amplitudes do
not seem to be directly linked to either the sign or amplitude of
$\Bo$) but certainly holds in an average sense (otherwise we would not
be able to observe regular field reversals).

Another possible (but as yet unscrutinized) explanation to explain the average
linear trend described above is that each new MRI-unstable shearing wave
packet involved in the chimera dynamics is nonlinearly seeded through
a physical process of scattering of its predecessor off the radially
modulated large-scale axisymmetric magnetic field (see H11 for a
detailed discussion of this effect), resuting in some statistical
correlation between the cumulated nonlinear EMF associated with 
swinging shearing waves and the large-scale field. The early
polarization of shearing wave seeds in their strongly leading phase, 
though, may not be particularly optimal in terms of MRI
amplification which, considering their limited lifetime,
may explain why they produce seemingly wildly different EMFs
(we recall that each shearing wave only has a finite time to grow
before it is strongly damped -- for $LB_{2m}$ with $L_y/L_x=4$ aspect
ratio, their ``active'' lifetime is of the order of few $\Omega^{-1}$
at most). Note that existing quasilinear statistical models randomly
drawing shearing wave seeds are currently blind to such scattering
effects. Whether this is a limitation remains to be assessed.

Implicit to the discussion above is that the sustainement of the
dynamo results from the cumulated effect of MRI-active 
\textcolor{black}{perturbations} as defined in Sect.~\ref{budget}. While it contrasts
with the classical view of mean-field effects being related to small-scale,
``inertial-range'' turbulence, this hypothesis is definitely supported
by our numerical simulations and magnetic energy budget analysis\footnote{To
  complicate the matters further, note that such linear
   growth effects are in principle captured by test field method
   calculations \citep{gressel10,gressel15}, whose original
   motivations are rooted in the classical mean field theory.}, as 
well as by other recent numerical results \citep{bhat16}. 
However, another result of the analysis conducted in Sect.~\ref{budget} is
 that smaller-scale slaved waves excited mostly through nonlinear
interactions (which are also not self-consistently computed in
quasi-linear models) are also important in this problem, and notably
play an important role in the magnetic Prandtl number dependence 
of the dynamo through the effect of turbulent magnetic dissipation.

Hence, it is tempting at this stage to sketch a refined physical
picture of the full turbulent MRI dynamo process (and more generally
of turbulent instability-driven dynamos) in which the EMF associated
with non-universal, fairly large-scale active modes feeling the effect
of the linear physics constitutes the main engine of the dynamo, while
faster, smaller-scale ``inertial-range'' slaved modes can either 
impede it (as in our chimera cycles) or reinforce it further (depending
on the problem). If this physical picture holds, then the decomposition
presented in Sect.~\ref{formalism} may prove extremely useful to
construct more transparent, physically-grounded statistical
mean-field dynamo models out of the existing ones.

\section{Conclusions}
\label{conclusions}
Motivated by the quest for a physically-grounded description of the
nonlinear process of turbulent dynamo action and
angular momentum transport in astrophysical accretion disks, we have
computed several new periodic, three-dimensional, fully nonlinear
incompressible magnetorotational dynamo solutions in moderate aspect
ratio shearing boxes. The dynamical complexity of these ``chimera''
solutions is significantly larger than that of solutions identified
earlier by H11, R13 and R15, and is reminiscent of the seemingly
complex statistical magnetic organization observed in many  fully
turbulent simulations of the problem. Yet, we have shown that their
sustainement can be understood in terms of the same few linear and
nonlinear dynamical processes underlying simpler cycles. These
solutions, like their simpler counterparts (R15), are also not
sustained for magnetic Prandtl numbers smaller than a few.

In order to understand the dynamics in a physically transparent way,
we have introduced a decomposition into active and slaved modes. The
former include a large-scale axisymmetric MRI-supporting field
component, and non-axisymmetric MRI-unstable energy-injecting
\textcolor{black}{perturbations}. The latter consist of 
\textcolor{black}{perturbations} passively
excited through nonlinear interactions that drain energy from larger
scales. Using this decomposition, we have been able to understand how
the magnetic energy of the system can be sustained via the MRI,
and to confirm the results of R15 regarding the role of turbulent
magnetic dissipation in the seeming disappearance of the MRI dynamo 
at low Pm. With this basic effect identified and confirmed, it may 
now be possible to better understand how it affects the dynamo and
turbulent  angular momentum transport in different geometric and
physical configurations \citep{shi16,walker16}.

The results presented in this paper are obviously not in the
astrophysically asymptotic regimes and do not accomodate all
\textcolor{black}{the relevant physics in this context}, such as magnetic 
buoyancy and stratification effects. However, we have shown that our 
physically transparent, fully three-dimensional, nonlinear 
magnetorotational dynamo chimeras
share some interesting properties with existing effective
two-dimensional statistical models of accretion disk dynamo cycles and
as such seem to offer an interesting path in parameter space towards
statistical asymptotic regimes. This raises the prospects that
improved effective statistical models of the MRI dynamo and other
instability-driven dynamos \citep{spruit02,cline03,miesch07,rincon08}
can be derived  from physical first principles, and may in the near
future provide trustable insights into magnetic field generation and
turbulent transport processes in a variety of stellar and circumstellar
environments.

\begin{acknowledgements}
We thank Richard Kerswell, S\'ebastien Fromang, Jonathan Squire and
Oliver Gressel for several useful discussions. This research was
supported by the University Paul Sabatier of Toulouse under an AO3
grant, by the Midi-Pyr\'en\'ees region, by the French National Program
for Stellar Physics (PNPS), by the Leverhulme Trust Network for
Magnetized Plasma Turbulence and by the National Science Foundation
under Grant No. PHY05-51164. Numerical calculations were carried out
on the CALMIP platform (CICT, University of Toulouse), whose
assistance is gratefully acknowledged.
\end{acknowledgements}



\bibliographystyle{aa}
\bibliography{refs} %if your bibtex file is called example.bib

\begin{thebibliography}{36}
\expandafter\ifx\csname natexlab\endcsname\relax\def\natexlab#1{#1}\fi

\bibitem[{{Balbus} \& {Hawley}(1991)}]{balbus91}
{Balbus}, S.~A. \& {Hawley}, J.~F. 1991, ApJ, 376, 214

\bibitem[{{Balbus} \& {Hawley}(1992)}]{balbus92}
{Balbus}, S.~A. \& {Hawley}, J.~F. 1992, ApJ, 400, 610

\bibitem[{{Balbus} \& {Henri}(2008)}]{balbus08}
{Balbus}, S.~A. \& {Henri}, P. 2008, ApJ, 674, 408

\bibitem[{{Bhat} {et~al.}(2016){Bhat}, {Ebrahimi}, \& {Blackman}}]{bhat16}
{Bhat}, P., {Ebrahimi}, F., \& {Blackman}, E.~G. 2016, submitted
  [arXiv:1605.02433]

\bibitem[{{Blackman}(2012)}]{blackman12}
{Blackman}, E.~G. 2012, Phys. Scr., 86, 058202

\bibitem[{{Brandenburg} {et~al.}(1995){Brandenburg}, {\hspace{-1pt}Nordlund},
  {\hspace{-0.5pt}Stein}, \& {\hspace{-1pt}Torkelsson}}]{branden95}
{Brandenburg}, {\hspace{-1pt}A}., {\hspace{-1pt}Nordlund}, A.,
  {\hspace{-0.5pt}Stein}, R.~{\hspace{-1pt}F}., \& {\hspace{-1pt}Torkelsson},
  {\hspace{-1pt}U}. 1995, ApJ, 446, 741

\bibitem[{{Brandenburg} \& {Subramanian}(2005)}]{branden05}
{Brandenburg}, A. \& {Subramanian}, K. 2005, Phys. Rep., 417, 1

\bibitem[{{C}handrasekhar(1960)}]{chandra60}
{C}handrasekhar, S. 1960, {Proc. Natl. Acad. Sci.}, 46, 253

\bibitem[{{Cline} {et~al.}(2003){Cline}, {Brummell}, \& {Cattaneo}}]{cline03}
{Cline}, K.~S., {Brummell}, N.~H., \& {Cattaneo}, F. 2003, ApJ, 599, 1449

\bibitem[{{Davis} {et~al.}(2010){Davis}, {Stone}, \& {Pessah}}]{davis10}
{Davis}, S.~W., {Stone}, J.~M., \& {Pessah}, M.~E. 2010, ApJ, 713, 52

\bibitem[{{Fromang} {et~al.}(2007){Fromang}, {Papaloizou}, {Lesur}, \&
  {Heinemann}}]{fromang07b}
{Fromang}, S., {Papaloizou}, J., {Lesur}, G., \& {Heinemann}, T. 2007, A{\&}A,
  476, 1123

\bibitem[{{Goldreich} \& {Lynden-Bell}(1965)}]{goldreich65}
{Goldreich}, P. \& {Lynden-Bell}, D. 1965, MNRAS, 130, 125

\bibitem[{{Gressel}(2010)}]{gressel10}
{Gressel}, O. 2010, MNRAS, 405, 41

\bibitem[{{Gressel} \& {Pessah}(2015)}]{gressel15}
{Gressel}, O. \& {Pessah}, M.~E. 2015, ApJ, 810, 59

\bibitem[{{Hawley} {et~al.}(1995){Hawley}, {Gammie}, \& {Balbus}}]{hawley95}
{Hawley}, J.~F., {Gammie}, C.~F., \& {Balbus}, S.~A. 1995, ApJ, 440, 742

\bibitem[{{Hawley} {et~al.}(1996){Hawley}, {Gammie}, \& {Balbus}}]{hawley96}
{Hawley}, J.~F., {Gammie}, C.~F., \& {Balbus}, S.~A. 1996, ApJ, 464, 690

\bibitem[{{Herault} {et~al.}(2011){Herault}, {Rincon}, {Cossu}, {Lesur},
  {Ogilvie}, \& {Longaretti}}]{Herault2011}
{Herault}, J., {Rincon}, F., {Cossu}, C., {et~al.} 2011, Phys. Rev. E, 84,
  036321

\bibitem[{{Lesur} \& {Longaretti}(2007)}]{lesur07}
{Lesur}, G. \& {Longaretti}, P.-Y. 2007, MNRAS, 378, 1471

\bibitem[{{Lesur} \& {Ogilvie}(2008{\natexlab{a}})}]{lesur08b}
{Lesur}, G. \& {Ogilvie}, G.~I. 2008{\natexlab{a}}, MNRAS, 391, 1437

\bibitem[{{Lesur} \& {Ogilvie}(2008{\natexlab{b}})}]{lesur08}
{Lesur}, G. \& {Ogilvie}, G.~I. 2008{\natexlab{b}}, A{\&}A, 488, 451

\bibitem[{{Meheut} {et~al.}(2015){Meheut}, {Fromang}, {Lesur}, {Joos}, \&
  {Longaretti}}]{meheut15}
{Meheut}, H., {Fromang}, S., {Lesur}, G., {Joos}, M., \& {Longaretti}, P.-Y.
  2015, AA, 579, A117

\bibitem[{{Miesch} {et~al.}(2007){Miesch}, {Gilman}, \& {Dikpati}}]{miesch07}
{Miesch}, M.~S., {Gilman}, P.~A., \& {Dikpati}, M. 2007, ApJ. Supp. Ser., 168,
  337

\bibitem[{{Moffatt}(1977)}]{moffatt77}
{Moffatt}, H.~K. 1977, {Magnetic field generation in electrically conducting
  fluids.} (Cambridge University Press)

\bibitem[{{Oishi} \& {Mac Low}(2011)}]{oishi11}
{Oishi}, J.~S. \& {Mac Low}, M.-M. 2011, ApJ, 740, 18

\bibitem[{{Rincon} {et~al.}(2007){Rincon}, {Ogilvie}, \& {Proctor}}]{rincon07b}
{Rincon}, F., {Ogilvie}, G.~I., \& {Proctor}, M.~R.~E. 2007, Phys. Rev. Lett.,
  98, 254502

\bibitem[{{Rincon} {et~al.}(2008){Rincon}, {Ogilvie}, {Proctor}, \&
  {Cossu}}]{rincon08}
{Rincon}, F., {Ogilvie}, G.~I., {Proctor}, M.~R.~E., \& {Cossu}, C. 2008,
  Astron. Nachr., 329, 750

\bibitem[{{Riols} {et~al.}(2015){Riols}, {Rincon}, {Cossu}, {Lesur}, {Ogilvie},
  \& {Longaretti}}]{riols15}
{Riols}, A., {Rincon}, F., {Cossu}, C., {et~al.} 2015, AA, 575, A14

\bibitem[{{Riols} {et~al.}(2013){Riols}, {Rincon}, {Cossu}, {Lesur}, {Ogilvie},
  {Longaretti}, \& {Herault}}]{riols2013}
{Riols}, A., {Rincon}, F., {Cossu}, C., {et~al.} 2013, J. Fluid Mech., 731, 1

\bibitem[{{Shi} {et~al.}(2016){Shi}, {Stone}, \& {Huang}}]{shi16}
{Shi}, J.-M., {Stone}, J.~M., \& {Huang}, C.~X. 2016, MNRAS, 456, 2273

\bibitem[{{Simon} {et~al.}(2011){Simon}, {Hawley}, \& {Beckwith}}]{simon11}
{Simon}, J.~B., {Hawley}, J.~F., \& {Beckwith}, K. 2011, ApJ, 730, 94

\bibitem[{{Spruit}(2002)}]{spruit02}
{Spruit}, H.~C. 2002, A{\&}A, 381, 923

\bibitem[{{Squire} \& {Bhattacharjee}(2015)}]{squire15}
{Squire}, J. \& {Bhattacharjee}, A. 2015, Phys. Rev. Lett., 114, 085002

\bibitem[{{Steenbeck} {et~al.}(1966){Steenbeck}, {Krause}, \&
  {R{\"a}dler}}]{steenbeck66}
{Steenbeck}, M., {Krause}, F., \& {R{\"a}dler}, K.-H. 1966, Z. Naturforschung
  Teil A, 21, 369

\bibitem[{{Stone} {et~al.}(1996){Stone}, {Hawley}, {Gammie}, \&
  {Balbus}}]{stone96}
{Stone}, J.~M., {Hawley}, J.~F., {Gammie}, C.~F., \& {Balbus}, S.~A. 1996, ApJ,
  463, 656

\bibitem[{{Velikhov}(1959)}]{velikhov59}
{Velikhov}, E.~P. 1959, Sov. Phys. JETP, 36, 1398

\bibitem[{{Walker} {et~al.}(2016){Walker}, {Lesur}, \& {Boldyrev}}]{walker16}
{Walker}, J., {Lesur}, G., \& {Boldyrev}, S. 2016, MNRAS, 457, L39

\end{thebibliography}

\end{document}